\documentclass[reprint,superscriptaddress,amsmath,amssymb,aps,pra,longbibliography
]{revtex4-1}

\usepackage[T1]{fontenc}
\usepackage[utf8]{inputenc}
\usepackage[unicode=true,breaklinks=true,colorlinks=true]{hyperref}
\hypersetup{citecolor=blue,urlcolor=blue}
\usepackage{graphicx}
\usepackage{bbm}
\usepackage{bbold}
\usepackage{lineno}

\usepackage{mathtools}
\usepackage{tensor}

\usepackage{pdfpages}
\usepackage{tikz}
\makeatletter
\AtBeginDocument{\let\LS@rot\@undefined}
\makeatother

\newcommand{\mh}[1]{\textcolor{black}{#1}}

\newcommand{\revise}[1]{\textcolor{black}{#1}}
\newcommand{\rev}[1]{\textcolor{black}{#1}}
\begin{document}
\title{Quantum Fisher information measurement and verification of the quantum Cram\'er-Rao bound in a solid-state qubit}
\author{Min Yu}
\thanks{These authors contributed equally.}
\author{Yu Liu}
\thanks{These authors contributed equally.}
\author{Pengcheng Yang}
\email{pengchengyang@hust.edu.cn}
\author{Musang Gong}
\affiliation{School of Physics, International Joint Laboratory on Quantum Sensing and Quantum Metrology, Institute for Quantum Science and Engineering, Huazhong University of Science and Technology, Wuhan 430074, China}
\author{Qingyun Cao}
\affiliation{School of Physics, International Joint Laboratory on Quantum Sensing and Quantum Metrology, Institute for Quantum Science and Engineering, Huazhong University of Science and Technology, Wuhan 430074, China}
\affiliation{	Institut f\"{u}r Quantenoptik $\&$ IQST, Albert-Einstein Allee 11, Universit\"{a}t Ulm, D-89081, Germany}
\author{Shaoliang Zhang}
\author{Haibin Liu}
\affiliation{School of Physics, International Joint Laboratory on Quantum Sensing and Quantum Metrology, Institute for Quantum Science and Engineering, Huazhong University of Science and Technology, Wuhan 430074, China}
\author{Markus Heyl}
\affiliation{Max Planck Institute for the Physics of Complex Systems, N{\"o}thnitzer Stra{\ss}e 38, Dresden 01187, Germany}
\author{Tomoki Ozawa}
\affiliation{Advanced Institute for Materials Research, Tohoku University, Sendai 980-8577, Japan}
\affiliation{Interdisciplinary Theoretical and Mathematical Sciences Program (iTHEMS), RIKEN, Wako, Saitama 351-0198, Japan}
\author{Nathan Goldman}
\email{ngoldman@ulb.ac.be}
\affiliation{Center for Nonlinear Phenomena and Complex Systems, Universit\'e Libre de Bruxelles, CP 231, Campus Plaine, B-1050 Brussels, Belgium}
\author{Jianming Cai}
\email{jianmingcai@hust.edu.cn}
\affiliation{School of Physics, International Joint Laboratory on Quantum Sensing and Quantum Metrology, Institute for Quantum Science and Engineering, Huazhong University of Science and Technology, Wuhan 430074, China}
\affiliation{State Key Laboratory of Precision Spectroscopy, East China Normal University, Shanghai, 200062, China}
\affiliation{Wuhan National Laboratory for Optoelectronics, Huazhong University of Science and Technology, Wuhan 430074, China}
\date{\today}
	
\begin{abstract}
The quantum Cram\'er-Rao bound sets a fundamental limit on the accuracy of unbiased parameter estimation in quantum systems, relating the uncertainty in determining a parameter to the inverse of the quantum Fisher information. We experimentally demonstrate near saturation of the quantum Cram\'er-Rao bound in the phase estimation of a solid-state spin system, provided by a nitrogen-vacancy center in diamond. This is achieved by comparing the experimental uncertainty in phase estimation with an independent measurement of the related quantum Fisher information. The latter is independently extracted from coherent dynamical responses of the system under weak parametric modulations, without performing any quantum-state tomography. \mh{While optimal parameter estimation has already been observed for quantum devices involving a limited number of degrees of freedom, our method offers a versatile and powerful \textit{experimental} tool to explore the Cram\'er-Rao bound and the quantum Fisher information in systems of higher complexity, as relevant for quantum technologies}.
\end{abstract}

\keywords{Quantum Fisher Information, Quantum Metrology, Quantum Sensing}
	
\maketitle

\noindent
{\bf INTRODUCTION}\\
Quantum metrology has emerged as a key quantum technological application. It allows for the improvement of sensors performance, beyond any classically achievable precision, as was demonstrated for instance in squeezed-light-based gravitational wave detectors~\cite{2011Ligo}. According to the quantum Cram\'er-Rao bound, the accuracy of any unbiased estimation of an unknown system parameter is limited by the inverse of the quantum Fisher information (QFI)~\cite{cramer1946,Rao1992,Braunstein1994,Braunstein1996,Petzbook2011,Toth2014,sidhu2020geometric}. Importantly, the QFI only depends on the quantum state and is independent of the estimator; it is a geometric property of a quantum state in parameter space. Thus, for each parameter estimation problem, there potentially exists an optimal quantum measurement that saturates the Cram\'er-Rao bound. Such fully efficient estimators can be found for classical systems and for small quantum devices upon comparing to theoretical predictions~\cite{Brida:2010} or by performing full-state tomography~\cite{Pan2019}, which, however, becomes extremely challenging for quantum systems with higher complexity. Consequently, the identification of optimal quantum measurement schemes would highly benefit from a universal method to measure the QFI within the experimental setting. In general, this is a complicated task~\cite{Strobel2014,Li2019,Pan2019}, which requires (by definition) a very precise determination of the "distance" (fidelity) between two quantum states. \revise{The quadratic coefficients of several fidelity-like quantities, such as Loschmidt echo~\cite{Macri2016}, Hellinger distance~\cite{Strobel2014,Li2019}, Euclidean distance~\cite{Zhang2017} and Bures distance~\cite{Frowis2016}, are related to the QFI. Hence, in principle, this allows for the evaluation of the QFI from the measurement of these quantities. The corresponding experiments have been demonstrated in an optical system~\cite{Zhang2017} and in Bose-Einstein condensates~\cite{Strobel2014}. In experiment, these quantities are usually determined by the statistical distances of two experimental probability distributions, which are obtained by measuring two quantum states upon an infinitesimally small change of the system parameters ~\cite{Braunstein1994,Braunstein1996,Pezze2016}. Considering these methods, the accurate estimation of the QFI requires precise control of system parameters and the ability to perform multiple measurements or even complete measurements~\cite{Frowis2016} on the system; this usually scales exponentially with the system size and remains challenging in many-qubit systems. Furthermore, the lower bound of the QFI can be obtained using quantum optimal control methods \cite{Yang2020}, variational algorithms \cite{Beckey2020,Meyer2021}, and random measurements \cite{Yu2021,Rath2021}, which typically require a large number of iterations or measurements.} 
In this work, we use a nitrogen-vacancy center in diamond to perform a fully efficient phase-estimation quantum measurement by showing saturation of the Cram\'er-Rao bound. In contrast to a previous study~\cite{Brida:2010}, where a saturation of the bound was identified through a theoretical estimation of the QFI, we \mh{hereby demonstrate saturation through purely experimental means by independently measuring the QFI} within our phase-estimation setting. \revise{This was achieved by directly probing spectroscopic responses upon weak parametric modulations, a technique which circumvents the stringent requirements of quantum-state tomography and avoids heavy experimental measurement overhead.} This has the advantage of offering a more scalable approach to more complex systems. Our method is inspired by a proposal to extract the quantum metric tensor~\cite{Ozawa2018,ozawa2019probing}, which was recently implemented in NV centers~\cite{Yu2018,chen2020experimental} and superconducting qubits~\cite{Tan2019}. We demonstrate this approach in a Ramsey interferometer, which represents a standard experimental setting for the estimation of an unknown phase parameter. We determine the optimal sensitivity of the phase-parameter estimation through different resource states, and compare these results with their individual QFI.
Finally, we demonstrate the applicability of our QFI measurement to the case of coupled qubits, and discuss its relation to entanglement signatures.\\

%
\begin{figure}
\hspace{-0.5cm}
\includegraphics[width=0.5\textwidth]{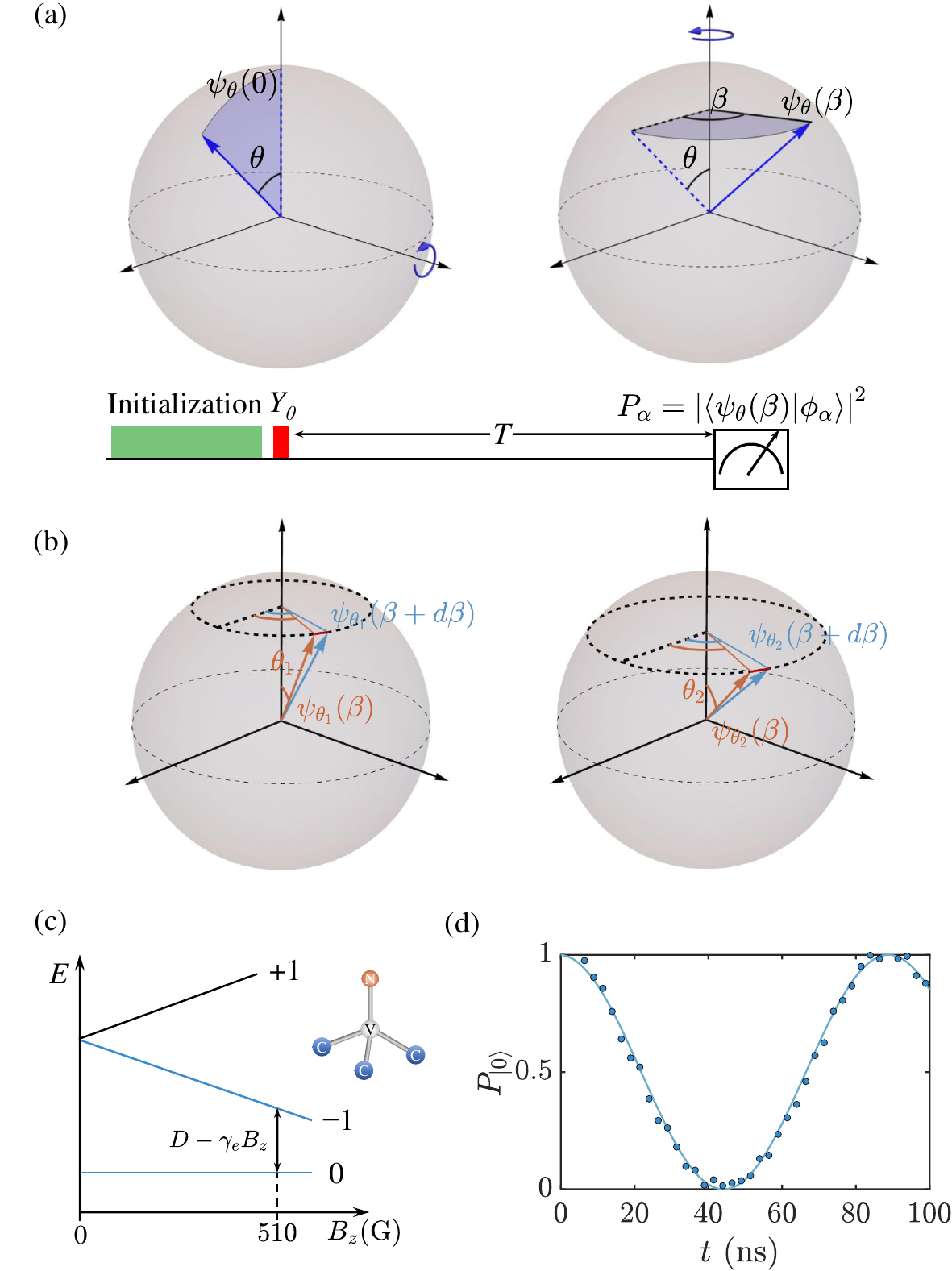}
\caption{{\bf Experimental setting.} {({\bf a})} Ramsey interferometry experiment for the estimation of an unknown phase parameter $\beta$. The quantum system is prepared in an initial resource state $|\psi_{\theta}(0)\rangle$, the evolution of which results in a phase parameter $\beta$. The projective measurement on the final state $|\psi_{\theta}(\beta)\rangle$ allows to determine the value of the parameter $\beta$. {({\bf b})} The QFI of the final state $|\psi_{\theta}(\beta)\rangle$ reveals the information content relative to the unknown phase parameter $\beta$. The larger QFI (right) implies the better distinguishability between the states $|\psi_{\theta}(\beta)\rangle$ and  $|\psi_{\theta}(\beta+d \beta)\rangle$ that have an infinitesimal parametric difference $d \beta\rightarrow 0$. {({\bf c})} The energy level structure of the NV center spin in diamond under an external magnetic field. The two-level quantum system is encoded by the ground state spin sublevels $m_s=0,-1$. {\bf ({\bf d})} Rabi oscillations:~the population in the spin state $m_s=0$ as a function of time, which facilitates efficient coherent control of the NV center spin state.}
\label{fig:setting}
\end{figure}
%


\noindent
{\bf RESULTS}\\
\noindent
{\bf Experimental setting}\\
%
In the experiment, we utilize a nitrogen-vacancy center (NV) in diamond as the quantum sensor.
The ground state of the NV center spin has three spin sublevels $m_s\!=\!\pm 1,0$. By applying an external magnetic field $B_z\simeq 510$ G along the NV axis, we lift the degeneracy of the spin states $m_s = \pm 1$ and use the two spin sublevels $m_s=0,-1$, with states $|0\rangle$ and $|-1\rangle$, to form a quantum two-level system with an energy gap $\omega_0=D-\gamma_e B_z$, where the zero-field splitting is $D=(2\pi)2.87$ GHz and $\gamma_e$ is the electronic gyromagnetic ratio [Fig.\ref{fig:setting}(c)]. We use a microwave field to coherently manipulate the NV center spin sate; see Fig.\ref{fig:setting}(d) for an illustrative Rabi oscillation.
%

%
\begin{figure*}
\centering
\includegraphics[width=0.83\textwidth]{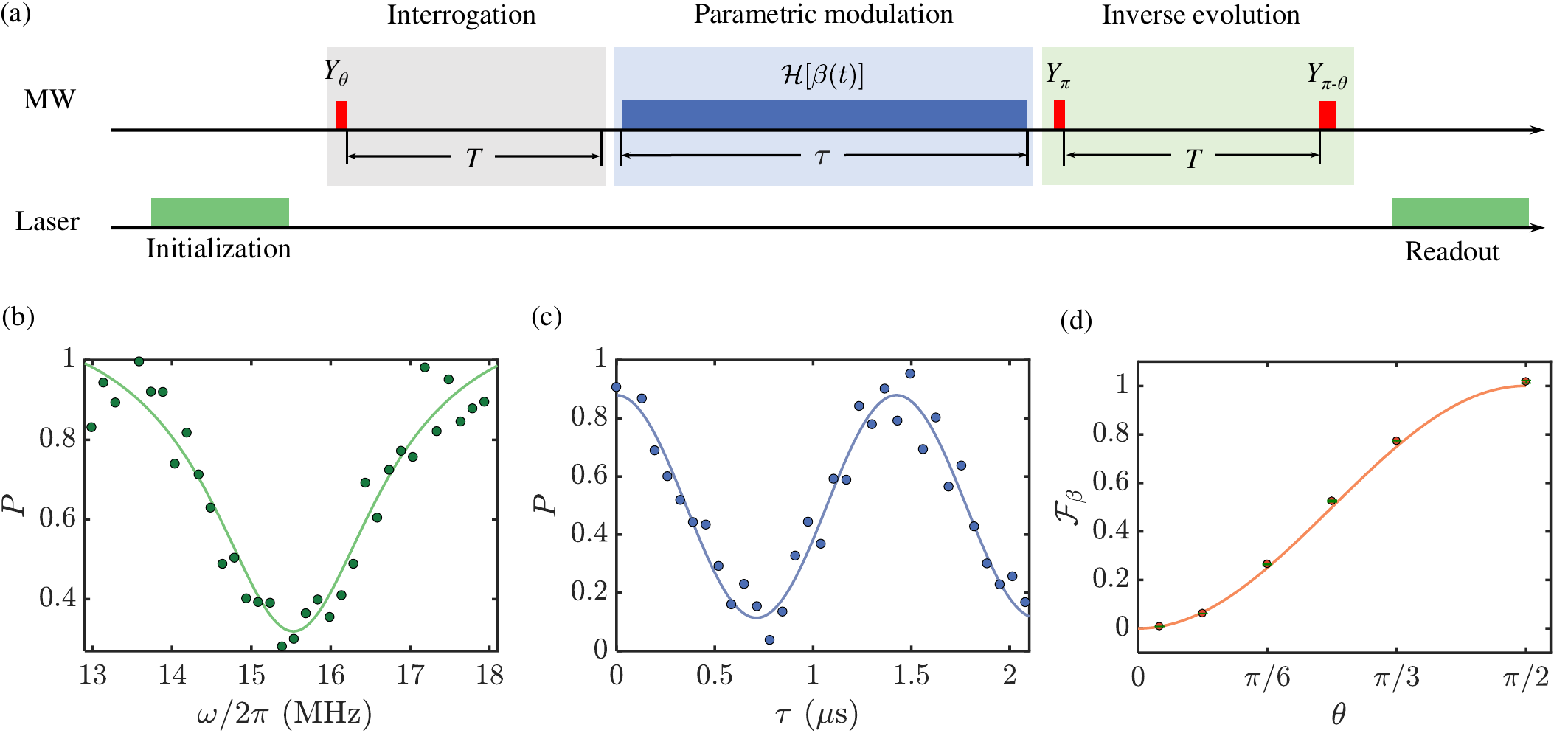}
\caption{\textbf{Direct measurement of the QFI.} ({\bf a}) The pulse sequence for the measurement of the QFI using the NV center spin. The  NV center spin is first polarized in the state $|0\rangle$ by applying a green (532 nm) laser pulse and the $\theta$-dependent resource state $|\psi_{\theta}(0)\rangle$ is prepared via a subsequent microwave pulse $Y_{\theta}$. The interrogation (i.e. the free evolution) for time $T$ results in the parameter-dependent final state $|\psi_{\theta}(\beta)\rangle$. The parametric modulation via the amplitude and phase modulated microwave driving is described by the Hamiltonian $H[\beta(t)]$ with $\beta(t)=\beta+a_{\beta}\cos(\omega t)$. The spin-dependent fluorescence after the inverse evolution, which rotates the state $|\psi_{\theta}(\beta)\rangle$ back to the state $|0\rangle$, monitors the coherent transition probability induced by the parametric modulation. ({\bf b}) The parameter-modulation induced resonant transition measurement shows the probability that the NV center spin stays in the state $|\psi_{\theta}(\beta)\rangle$ as a function of the modulation frequency $\omega$ for a time $\tau=450$ ns. ({\bf c}) The resonant coherent oscillation between the state $|\psi_{\theta}(\beta)\rangle$ and $|\psi^{\perp}_{\theta}(\beta)\rangle$ under parametric modulation. The other experiment parameters in ({\bf b}) and ({\bf c}) are $\theta=\pi/3$, $A=(2\pi)15.98$ MHz, $a_{\beta}=0.1$ and $\revise{\xi}=(2\pi)5.025$ MHz. ({\bf d}) The QFI measured in our experiment (red circle) is compared with the theoretical prediction (brown curve).}
\label{fig:2}
\end{figure*}

%
\mh{Quantum sensing and parameter estimation have been implemented in NV centers using different approaches~\cite{Rondin2014, Degen2017}, inspired by the pioneer proposal and demonstration of magnetometry based on Ramsey spectroscopy~\cite{Taylor2008, Maze2008, Balasubramanian2008}. Building on those developments, we hereby adopt the standard protocol of a phase-parameter estimation measurement by means of Ramsey interferometry [Fig.\ref{fig:setting}(a)].} For that purpose, we first initialize the system in a coherent superposition resource state, $|\psi_{\theta}(0)\rangle=\cos{(\theta/2)}|0\rangle-\sin{(\theta/2)}|\!-1\rangle$,
which we evolve into
 \begin{equation}
|\psi_{\theta}(\beta)\rangle=\cos{(\theta/2)}e^{i\beta/2}|0\rangle-\sin{(\theta/2)}e^{-i\beta/2}|-1\rangle\, ,
\label{eq:final_state}
 \end{equation}
according to the applied magnetic field.
The phase parameter $\beta$ of $|\psi_{\theta}(\beta)\rangle$ can be estimated by performing positive-operator valued measurements (POVM)~\cite{Braunstein1996,sidhu2020geometric}, $\mathcal{M}=\{\mathcal{M}_j\}$; as explained below, these are provided by spin-dependent fluorescence measurements (see Appendix).
The measurement precision is defined as the minimal change of the parameter $\beta$ that can be detected from the constructed observable above the shot-noise level,
\begin{equation}
( \delta \beta)_{\mathcal{M}} = \Delta p /(\frac{\partial p}{\partial \beta}),\label{CRbound}
\end{equation}
where $p$ is the expectation value of the POVM signal, $\Delta p $ is the uncertainty associated with the measurement signal. The fundamental limit of the achievable sensitivity of an unbiased estimator is given by the quantum Cram\'er-Rao bound~\cite{Helstrom1976,Holevo2011,Hayashi2017}
\begin{equation}
\delta \beta\geq\frac{1}{\sqrt{\mathcal{F}_{\beta}}}\, ,
\label{eq.2:bound}
\end{equation}
where $\mathcal{F}_\beta$ denotes the QFI, which for pure quantum states $|\psi_{\theta}(\beta)\rangle$, is given by~\cite{Braunstein1994,Braunstein1996}
\begin{equation}
\mathcal{F}_{\beta}=4\left[\langle\partial_{\beta}\psi_{\theta}(\beta)|\partial_{\beta}\psi_{\theta}(\beta)\rangle-|\langle\psi_{\theta}(\beta)|\partial_{\beta}\psi_{\theta}(\beta)\rangle|^2\right].\label{Fisher}
\end{equation}
The QFI characterizes the distinguishability of adjacent quantum states over the parameter space [Fig.\ref{fig:setting}(b)]. The purity of the states in our experiment, and hence the validity of Eq.\eqref{Fisher} to capture the QFI, is discussed below. We note that the QFI is related to the real part of the quantum geometric tensor,  which can be extracted through coherent dynamical responses~\cite{Ozawa2018,Yu2018}.
%

%
 \begin{figure}
 \hspace{-0.5cm}
\includegraphics[width=0.5\textwidth]{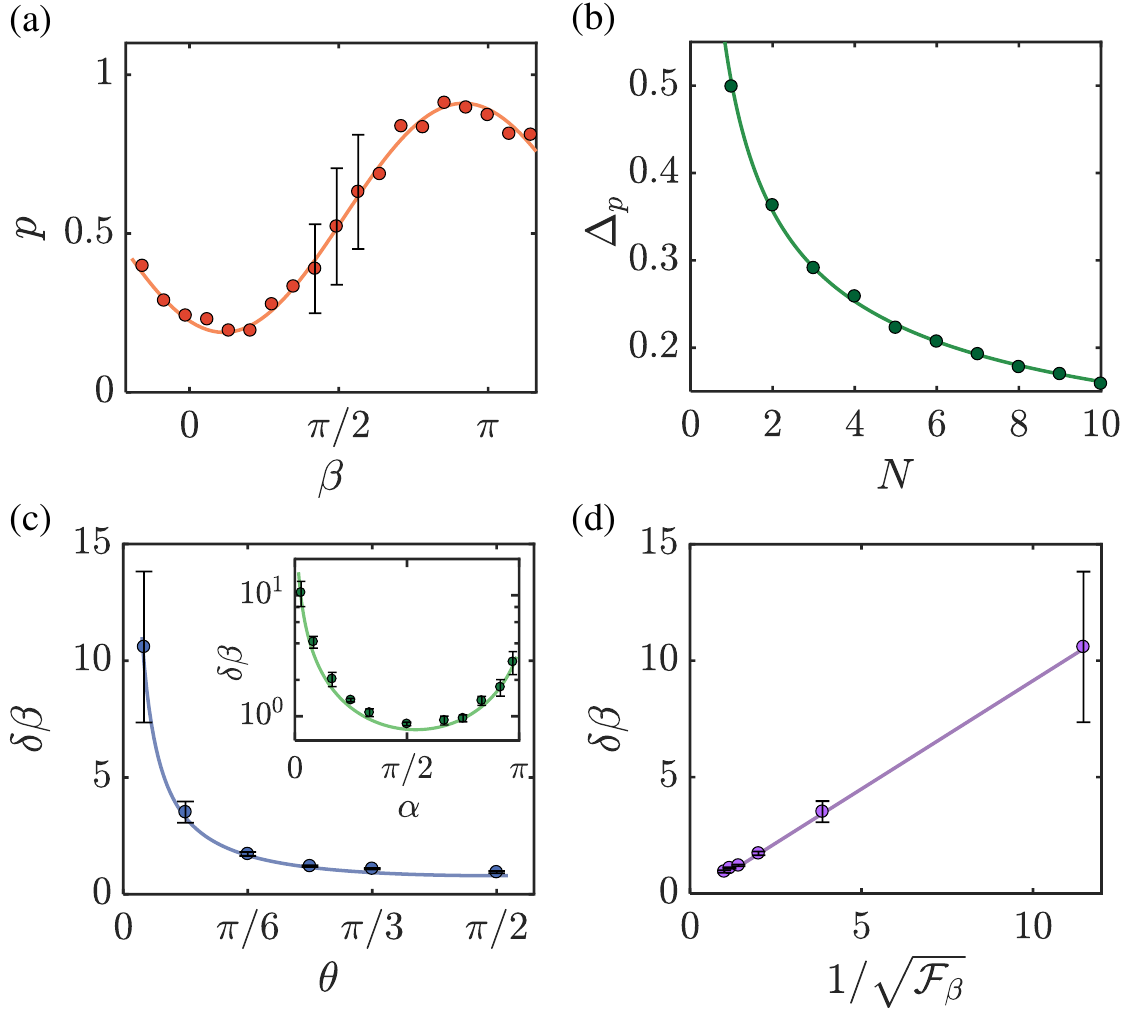}
 \caption{{\bf Saturating the quantum Cram\'er-Rao bound}. ({\bf a}) The Ramsey interferometry measurement signal $p\!=\!\langle S \rangle$. The measurement data allows us to obtain the susceptibility $\chi_{\alpha}=\partial p /\partial \beta$ of the measurement signal close to the working point $\beta=\pi/2$. The error bars represent the uncertainty of the parameter estimation $\Delta p=[\langle S^2\rangle -\langle S\rangle ^2]^{1/2}$  with the number of repetitions $N=9$. The parameters are $\theta=\pi/3$, $\alpha=\pi/2$, $\revise{\xi}=(2\pi)2.27$ MHz and $A=(2\pi)11.34$ MHz. ({\bf b}) The uncertainty of the parameter estimation $\Delta s$ as a function of the number of repetitions $N$ can be fitted by a function of the form $\Delta p= \Delta_0/\sqrt{N}$ (green curve). ({\bf c}) The optimal measurement sensitivity $\delta\beta$ (achieved by the projective measurement $P_{\alpha}$ with $\alpha=\pi/2$) by using different $\theta$-dependent resource states $\vert \psi_{\theta}(0)\rangle$. Inset: The sensitivity $\delta\beta$, achieved by applying the projective measurement $P_{\alpha}$ as a function of $\alpha$ when $\theta=\pi/2$ and $\beta=\pi/2$, shows that the optimal measurement sensitivity in our Ramsey interferometry experiment is obtained when $\alpha=\pi/2$. \mh{The green curve is obtained from numerical simulation, see further details in \revise{Fig.S4}.} ({\bf d}) The linear relation $\delta\beta\propto1/\sqrt{\mathcal{F}_{\beta}}$, where $\mathcal{F}_{\beta}$ is the quantum Fisher information; the measured proportionality factor is $1.041\pm0.036$. The number of repetitions in ({\bf c-d}) is $N=1$. The curves in ({\bf a, c-d}) are theoretical predictions.
 }
\label{fig:sensitivity}
\end{figure}
%

%

It is one of the central goals of this work to show the saturation of the quantum Cram\'er-Rao bound through an independent experimental measurement of the QFI.
We extract the QFI by probing coherent dynamical responses of the quantum system upon perturbative parametric modulations~\cite{Ozawa2018,Yu2018}. Our measurement protocol is shown in Fig.\ref{fig:2}(a).
The NV center spin is first initialized in the $m_s\!=\!0$ spin state by applying a green (532 nm) laser pulse, which also polarizes the nitrogen nuclear spin associated with the NV center as we tune the magnetic field close to the excited state level anticrossing (i.e.~$B_z \simeq 510$ Gauss). The subsequent microwave pulse, applied for a duration $t_{\theta}\!=\!(\theta/\Omega)$, rotates the NV center spin around the $\hat{y}$ axis by an angle $\theta$ according to the Hamiltonian $H_1(t)\!=\!(\omega_1/2)\sigma_z+\Omega\cos{(\omega_1 t)}\sigma_x$, where $\omega_1$ matches the energy gap between the spin sublevels $m_s\!=\!0, -1$ and $\Omega$ is the microwave Rabi frequency. The rotation, denoted as $Y_{\theta}$, prepares the NV center spin into the $\theta$-dependent resource state $|\psi_{\theta}(0)\rangle$. After the microwave pulse $Y_{\theta}$, the system undergoes a free evolution for a time  $T$, according to an effective Hamiltonian $H_{2}^{(e)}\!=\![(\omega_0-\omega_1)/2]\sigma_z$, which results in the final state $|\psi_{\theta}(\beta)\rangle$; see Eq.\eqref{eq:final_state}. Here, the effective Hamiltonian $H_{2}^{(e)}$ is defined in the interaction picture with respect to $H_0=(\omega_1/2)\sigma_z$. \revise{The final state $|\psi_{\theta}(\beta)\rangle$ encodes the information about the phase parameter $\beta\!=\!\xi T$ to be estimated, where $\xi\!=\!\omega_1-\omega_0$.}\\

\noindent
{\bf Direct measurement of the QFI}\\
Inspired by the protocol of Ref.~\cite{Ozawa2018}, we extract the QFI of the final state $|\psi_{\theta}(\beta)\rangle$ by monitoring coherent transitions upon parametric modulations. This probing method requires the implementation of the following Hamiltonian
    \begin{equation}
        \mathcal{H}(\beta)=\frac{A}{2}\left(\begin{array}{cc}
        \cos{\theta}                &  \sin{\theta}e^{-i\beta}  \\
         \sin{\theta}e^{i\beta}    &    -\cos{\theta}
        \end{array}\right),
        \label{eq:Hamiltonian}
    \end{equation}
such that the state $|\psi_{\theta}(\beta)\rangle$ approximately corresponds to an eigenstate of $\mathcal{H}(\beta)$. This is achieved by tuning the parameters of the microwave driving field acting on the NV center spin. The key step of our experiment then consists in generating parametric modulations~\cite{Ozawa2018}. \revise{To achieve this, we synthesize and calibrate an appropriate microwave driving field with proper amplitude and phase modulations, see Supplementary Note 1 and ~\cite{Yu2018,Ozawa2018} using an arbitrary waveform generator as follows
\begin{equation}
f_0(t)=(A\sin{\theta})\cos{[(\omega_1-A\cos{\theta})t+\beta(t)]},
\end{equation}
such that the "probing" Hamiltonian retains the form in Eq.~\eqref{eq:Hamiltonian}, but with a time-periodic modulation of the parameter $\beta$, i.e.~$\mathcal{H}(\beta)\!\rightarrow\!\mathcal{H}[\beta(t)]=\mathcal{H}(\beta+a_{\beta}\cos(\omega t))$, where $a_{\beta}\ll 1$ quantifies the modulation amplitude.}
The parametric modulation can induce a coherent transition from the state $|\psi_{\theta}(\beta)\rangle$ to the other orthogonal eigenstate $|\psi_{\theta}^{\perp}(\beta)\rangle$ of the Hamiltonian in Eq.\eqref{eq:Hamiltonian} \cite{Ozawa2018,Yu2018}. This transition can be monitored by measuring the probability that the system remains in the state $|\psi_{\theta}(\beta)\rangle$. In the experiment, without requiring any prior information on the parameter $\beta$, we implement an inverse evolution sequence, consisting of two pulses ($Y_{\pi}$ and $Y_{\pi-\theta}$) separated by a free evolution of duration $T$ [Fig.\ref{fig:2}(a)]. Such an inverse evolution rotates the states $|\psi_{\theta}(\beta)\rangle$ and $|\psi_{\theta}^{\perp}(\beta)\rangle$ back to the states $|0\rangle$ and $|-1\rangle$, respectively, see Supplementary Note 1. We then measure the population in state $|0\rangle$, which equals to the sought population in state $|\psi_{\theta}(\beta)\rangle$ after the application of the parametric modulation.
The efficiency of the coherent transition induced by the modulation is optimal whenever the modulation frequency matches the energy gap between the states $|\psi_{\theta}(\beta)\rangle$ and $|\psi_{\theta}^{\perp}(\beta)\rangle$. In the experiment, we first perform the modulation-induced-transition measurement for a wide range of modulation frequencies, from which we determine the resonant modulation frequency $\omega\simeq A$; see Fig.\ref{fig:2}(b). We then apply the parametric modulation at the resonant frequency, and measure the population in the state  $|\psi_{\theta}(\beta)\rangle$ as a function of the perturbation duration $\tau$; see Fig.\ref{fig:2}(c). This data is fitted using a function $P_0\!=\![1+\cos{(\nu_{\theta} t)}]/2$, which defines the effective Rabi frequency $\nu_{\theta}$.  From this data, we extract the $\theta$-dependent QFI, $\mathcal{F}_{\beta}(\theta)$, using the relation (see Appendix)
\begin{equation}
\mathcal{F}_{\beta}(\theta)=4\left(\frac{\nu_{\theta}}{a_{\beta}\omega}\right)^2.
\end{equation}
This experimental measurement of the QFI is displayed in Fig.\ref{fig:2}(d), which shows excellent agreement with the theoretical prediction $\mathcal{F}_{\beta}\!=\!\sin^2{\theta}$. In particular, it clearly demonstrates the dependence of the QFI on the initial resource state $\vert \psi_{\theta}(0)\rangle$. The precision of our measurement relies on the accuracy of the engineered Hamiltonian $\mathcal H(\beta)$ and on the determination of the effective Rabi frequency $\nu_{\theta}$. The imperfection in the interrogation step [Fig.~\ref{fig:2}(a)] may result in a mixed state rather than a pure state $\vert \psi_{\theta}(\beta)\rangle$; this would decrease the contrast of the Rabi oscillations and affect the measurement accuracy. By reconstructing the density matrix through projective measurements, we estimate the state fidelity to be above $95\%$ in our experiment, see Supplementary Note 2, which is evidenced by the good agreement between our results and the theoretical predictions.\\

\noindent
{\bf Reaching the quantum Cram\'er-Rao bound}\\
%
%
%
The QFI measurement enables us to experimentally show that our phase-parameter estimator exhibits optimal performance by saturating the quantum Cram\'er-Rao bound in Eq.\eqref{eq.2:bound}. In order to analyze the relation between the measurement precision and the QFI, we now determine the measurement sensitivity for the estimation of the parameter $\beta$ within our Ramsey interferometry experiment. To do so, we first apply the rotation $Y_{\theta}$ on the NV center spin qubit to prepare the initial state $|\psi_{\theta} (0)\rangle$; the system then evolves freely for a time $T=\beta/\revise{\xi}$. To build an estimator of the parameter $\beta$, we apply a rotation $Y_{\alpha}$, which is equivalent to a projective measurement $P_{\alpha}\!=\!|\phi_{\alpha}\rangle\langle\phi_{\alpha}|$ on the final state $|\psi_{\theta}(\beta)\rangle$, where $|\phi_{\alpha}\rangle=\cos(\alpha/2)|0\rangle+\sin(\alpha/2)|-1\rangle$~[Fig.\ref{fig:setting}(a)]. The observable of interest is then provided by the function $p(\beta ; \theta, \alpha)\!=\!\langle \psi_{\theta}(\beta) \vert P_{\alpha}\vert \psi_{\theta}(\beta)\rangle$, from which we aim to estimate the parameter $\beta$ with optimal accuracy [Eq.\eqref{eq.2:bound}]. We tune the free evolution time such that the parameter $\beta=\revise{\xi} T$ is close to the working point where the best sensitivity occurs, i.e. $\beta\simeq\pi/2$ where the slope $\partial p/\partial\beta$ is maximal~[Fig.\ref{fig:sensitivity}(a)].
\mh{Ramsey parameter estimation can, in principle, achieve optimal efficiency. However, in practice, this would require an ideal projective measurement of the sensor upon reaching the shot-noise limit. Such an ideal measurement cannot be perfectly performed, due to a limited collection efficiency or other types of measurement noise (e.g. Gaussian fluctuations in the photon number). To overcome this limitation, one may adopt the technique of single-shot readout \cite{Neumann2010,Robledo2011,Dreau2013,Liu2017}, which consists in setting a threshold $n_s$ of photon number to distinguish the state $\vert m_s=-1\rangle$ and $\vert m_s=0\rangle$ and assign a value $s=0$ or $1$ depending on whether $n_j>n_s$ or $n_j<n_s$.} 
In our experiment, the \rev{observable $p(\beta ; \theta, \alpha)\!=\!\langle \psi_{\theta}(\beta) \vert P_{\alpha}\vert \psi_{\theta}(\beta)\rangle$ is estimated from the collected photons of a fluorescence signal} (see Appendix). Due to the limited collection efficiency, the signal photons are accumulated over many sweeps of an experimental  sequence, which constitutes one experimental run of our measurement. \revise{In the $j$-th  run, based on the photon number $n_j$ detected from the rotated spin state $Y_{\alpha} |\psi_{\theta}(\beta)\rangle$, we define the ratio  $p_j=(n_j-n_1)/(n_0-n_1)$ where $n_0$ and $n_1$ are the average photon numbers obtained from the bare spin states $m_s\!=\!0$ and $m_s\!=\!-1$, respectively. We proceed to assign a measurement value $s_j=k+1$ or $k$ according to the probabilities $p_j^{(k)}=p_j-k $ and $1-p_j^{(k)}$ for $\lfloor p_j\rfloor=k$, see Supplementary Note 2. This allows us to introduce a quantity $S\!=\!(1/N)\sum_{j=1}^N s_j$, whose expectation value yields the desired function $\langle S \rangle\!=\!p(\beta ; \theta, \alpha)$. Using this quantity, we can construct an estimator for the parameter $\beta$, and find that the influence of measurement noise on $S$ is \rev{eliminated to a large extent} (apart from the shot-noise), which also provides a data analysis alternative for the spin readout techniques of NV centers~\cite{Neumann2010,Robledo2011,Dreau2013,Liu2017}, see Supplementary Note 2.} The data obtained from repeated measurements [Fig.\ref{fig:sensitivity}(a)] allows us to determine the slope of the signal, which is defined as $\chi_{\alpha}\!=\!\partial p /\partial \beta\!=\!\left[ p(\beta+d\beta)-p(\beta)\right ] /d\beta$. From the experimental data, we can also extract the measurement uncertainty $\Delta p$ associated with the observable $S$; see Fig.\ref{fig:sensitivity}(b). We note that the uncertainty scales with the number of repetitions $N$ as $\Delta p\!=\!\Delta_0 /\sqrt{N} + \xi_0 $, see Supplementary Note 2. The first term arises from the shot-noise with $\Delta_0\!=\![p(1-p)]^{1/2}$, while the second term $\xi_0$ represents the contribution from the measurement fluctuation that cannot be averaged out. \rev{We remark that other advanced readout techniques, such as the single-shot measurement based on spin to charge conversion \cite{Zhang2021}, can further reduce such measurement noise (see Eq.S29-S.30 in Supplementary Note 2) and enhance the sensitivity.}
We first compare the sensitivity $\delta \beta \!=\!\Delta p/ \chi_{\alpha}$ obtained by projective measurements over different bases $P_{\alpha}$. The experimental results shown in the inset of Fig.\ref{fig:sensitivity}(c) demonstrate that the optimal measurement sensitivity is obtained when $\alpha\!=\!\pi/2$, which agrees with the theoretical prediction (see Appendix), $(\delta\beta)^2\!=\![1-(\cos\beta \sin\theta)^2]/|\sin\beta \sin\theta|^2$. The slight deviation arises from other sources (apart from shot noise). The measurement precision also depends on the angle $\theta$ of the resource state $\vert \psi_{\theta}(0)\rangle$, which accounts for the QFI of the final state $\vert \psi_{\theta}(\beta)\rangle$:~we proceed by determining the optimal measurement sensitivity with different resource states $\vert \psi_{\theta}(0)\rangle$ in view of testing the quantum Cram\'er-Rao bound in Eq.\eqref{eq.2:bound}. It can be seen from the results shown in Fig.\ref{fig:sensitivity}(c) that the optimal measurement sensitivity improves as the angle $\theta$ approaches $\pi/2$, i.e. when the resource state $\vert \psi_{\theta}(0)\rangle$ becomes a maximally coherent superposition state. \mh{We remark that the result in the inset of Fig.\ref{fig:sensitivity}(c) is skewed as the pulse $Y_{\alpha}$ is off-resonant; the influence of the corresponding detuning is the asymmetry observed around $\alpha\!=\!\pi/2$.} Moreover, the optimal measurement sensitivity verifies the quantum Cram\'er-Rao bound~[Eq.\eqref{eq.2:bound}], as we finally demonstrated in Fig.\ref{fig:sensitivity}(d).\\

\noindent
{\bf Generalization to entangled qubits}\\
%
%
%
%
Single NV centers in diamond allow to perform quantum sensing with unprecedented spatial resolution~\cite{Grinolds2014}. In this context, the saturation of the Cram\'er-Rao bound is of particular importance as it may allow quantum sensing with unparalleled accuracy. Still it is a natural question whether our QFI measurement can also be extended to the multi-qubit case, where quantum entanglement can provide a further key factor to increase the performance of a quantum sensor. 

For that purpose, we now demonstrate the applicability of our parametric modulation scheme in view of measuring the QFI in a realistic two-qubit correlated system~\cite{Yu2018}, which consists of an NV center and a nearby strongly coupled $^{13}\mbox{C}$ nuclear spin via the hyperfine interaction. The effective Hamiltonian of the system is given by (see Supplementary Note 3)
\begin{eqnarray}
	\nonumber\mathcal{H}=&&\frac{A}{2}\left[\cos\beta\sigma_z+\sin\beta(\cos\phi\sigma_x+\sin\phi\sigma_y)\right]\\
	\nonumber&&-\frac{A_{\parallel}}{4}\sigma_z\tau_z-\frac{A_{\perp}}{4}\sigma_z\tau_x+\left(\frac{\omega_C}{2}-\frac{A_{\parallel}}{4}\right)\tau_z-\frac{A_{\perp}}{4}\tau_x,\\
\end{eqnarray}
where $\sigma$ and $\tau$ denote the Pauli matrices of the NV center and of the $^{13}\mbox{C}$ nuclear spin, respectively. We denote the four eigenstates of this Hamiltonian as $|\Psi_1\rangle$, $|\Psi_3\rangle$, $|\Psi_3\rangle$ and $|\Psi_4\rangle$, with their associated eigenvalues $\epsilon_1<\epsilon_2<\epsilon_3<\epsilon_4$. Similarly to the single qubit case treated above, we are interested in the quantum-parameter-estimation problem associated with the parameter $\beta$, and in particular, to the related QFI. Without loss of generality, we focus our study on the QFI contained in the lowest-energy eigenstate $|\Psi_1\rangle$.

\begin{figure}
	\hspace{-0.5cm}
\includegraphics[width=0.5\textwidth]{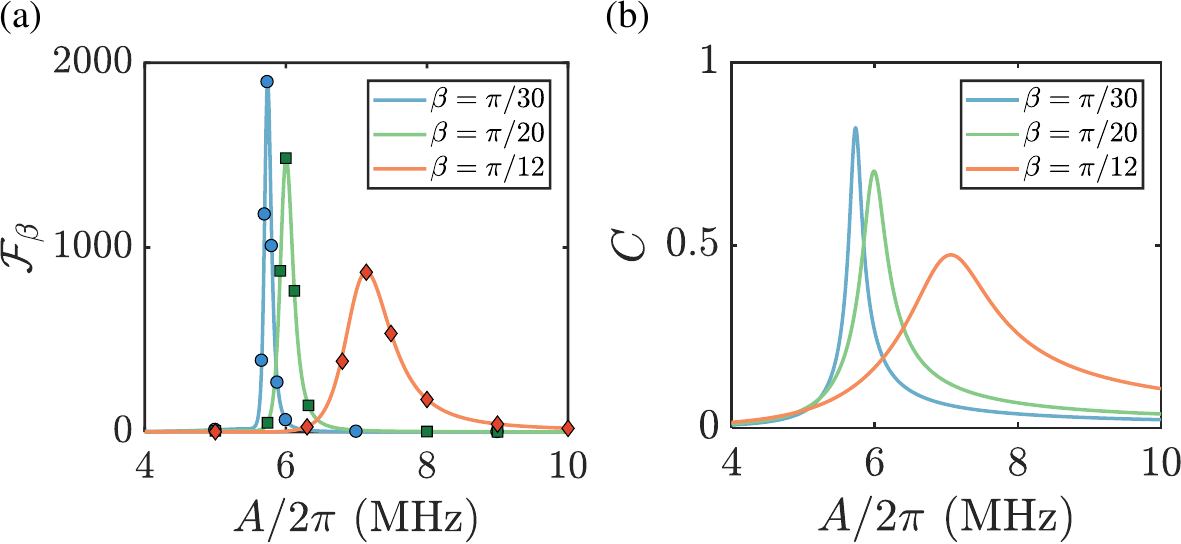}
	\caption{{\bf \mh{Numerical simulation of the QFI and quantum entanglement} in a strongly correlated two-qubit system}. ({\bf a}) The QFI $\mathcal{F}_{\beta}$ of the ground state $|\Psi_1\rangle$. The data points obtained from the simulation of the experiments agree well with the exact theoretical values (solid lines).  ({\bf b}) The concurrence $C$ of ground state $|\Psi_1\rangle$. The parameters we use are $A_{\perp}=(2\pi)2.79$ MHz, $A_{\parallel}=(2\pi)11.832$ MHz and $B_z=504$ G. The values of the modulation strength, $a_k$, are chosen such that the Rabi frequency of the induced coherent oscillation is much smaller than the corresponding energy gaps, see {Ref}.~\cite{Ozawa2018}.}
	\label{fig:entanglement}%
\end{figure}

%
Considering the parametric modulation $\beta(t)=\beta+a\cos(\omega t)$, the QFI can be related to the three Rabi frequencies $\nu_k$ associated with the induced transitions between the ground state $|\Psi_1\rangle$ and the other three eigenstates $|\Psi_k\rangle$ according to
\begin{equation}
	\mathcal{F}_{\beta}=4\sum_{k=2}^{4}\left(\frac{\nu_k}{a_k\omega_k}\right)^2,
\end{equation}
where $\omega_k=\epsilon_k-\epsilon_1$. We have performed a numerical simulation of this setting and we present the results in Fig.\ref{fig:entanglement}(a). We find that the QFI of the ground state reaches its peak value when the energy of the corresponding eigenstate becomes very close to another eigenenergy, in the form of an avoided crossing (see Appendix). In this situation, a small variation of the parameter (i.e. a perturbation) would indeed result in a significant change of the ground state. Importantly, this increase of the QFI is accompanied with a significant growth of entanglement, as quantified by the concurrence~\cite{Wootters1998}, as we demonstrate in Fig.\ref{fig:entanglement}(b). The connection between the QFI and the entanglement of such a coupled-qubit setting (see Appendix) is known to arise from the level anticrossing~\cite{Berkley2003,Smirnov2013}, which represents a general feature in systems beyond the single-qubit context. These results suggest that a large QFI is linked to strong entanglement upon measuring the QFI based on parametric modulations as introduced here. \revise{We remark that the proposed protocol can be extended to experimentally determine the QFI of many-body quantum systems by measuring the excitation rate under parametric modulation following the idea as presented in Ref. \cite{Ozawa2018}. The approach does not require full state tomography, which is an experimentally demanding task for a multi-qubit system. The present technique which allows us to estimate the QFI, and hence the quantum Cram\'er-Rao bound, will be helpful in resolving the challenging task of determining the optimal measurement for many-body ground states that can reach the bound.} \\

\noindent
{\bf DISCUSSION} \\
\mh{In this work, we have introduced an experimental technique to measure the QFI in a solid-state spin system based on spectroscopic responses. Importantly, this approach does not require full state tomography, and it can therefore be potentially applied to more complex systems. We have shown that this technique offers a genuine experimental probe of the quantum Cram\'er-Rao bound saturation, which does not rely on any theoretical knowledge, hence providing a universal tool to identify fully efficient estimators.} The presented technique provides a versatile tool to explore the fundamental role of the QFI in various physical scenarios, including quantum metrology, but also entanglement properties of many-body quantum systems~\cite{Hauke2016,ozawa2019probing} and the quantum speed limit in the context of optimal control~\cite{Giovannetti2003,Taddei2013,Campo2013,Pires2016,kolodrubetz2017geometry}.\\

\vspace{0.2cm}
\noindent
{\bf ACKNOWLEDGEMENTS} \\
This work is supported by the National Natural Science Foundation of China (11874024, 11690032), the National Key R$\&$D Program of China (Grant No. 2018YFA0306600), the Open Project Program of Wuhan National Laboratory for Optoelectronics (No. 2019WNLOKF002). T.O. is supported by JSPS KAKENHI Grant Number JP18H05857, JST PRESTO Grant Number JPMJPR19L2, JST CREST Grant Number JPMJCR19T1, and the Interdisciplinary Theoretical and Mathematical Sciences Program (iTHEMS) at RIKEN. N.G. is supported by the ERC Starting Grant TopoCold and the Fonds De La Recherche Scientifique (FRS-FNRS) (Belgium). This project has received funding from the European Research Council (ERC) under the European Union’s Horizon 2020 research and innovation programme (grant agreement No. 853443), and M.H. further acknowledges support by the Deutsche Forschungsgemeinschaft via the Gottfried Wilhelm Leibniz Prize program.\\


\onecolumngrid
 

\section*{Supplementary material}

# Supplementary Material for "Quantum Fisher information measurement and verification of the quantum Cramér-Rao bound in a solid-state qubit"

Min Yu,[1, *] Yu Liu,[1, *] Pengcheng Yang,[1, †] Musang Gong,[1] Qingyun Cao,[1, 2] Shaoliang Zhang,[1] Haibin Liu,[1] Markus Heyl,[3] Tomoki Ozawa,[4, 5] Nathan Goldman,[6, ‡] and Jianming Cai[1, 7, 8, §]

[1]*School of Physics, International Joint Laboratory on Quantum Sensing and Quantum Metrology, Institute for Quantum Science and Engineering, Huazhong University of Science and Technology, Wuhan 430074, China*
[2]*Institut für Quantenoptik & IQST, Albert-Einstein Allee 11, Universität Ulm, D-89081, Germany*
[3]*Max Planck Institute for the Physics of Complex Systems, Nöthnitzer Straße 38, Dresden 01187, Germany*
[4]*Advanced Institute for Materials Research, Tohoku University, Sendai 980-8577, Japan*
[5]*Interdisciplinary Theoretical and Mathematical Sciences Program (iTHEMS), RIKEN, Wako, Saitama 351-0198, Japan*
[6]*Center for Nonlinear Phenomena and Complex Systems, Université Libre de Bruxelles, CP 231, Campus Plaine, B-1050 Brussels, Belgium*
[7]*State Key Laboratory of Precision Spectroscopy, East China Normal University, Shanghai, 200062, China*
[8]*Wuhan National Laboratory for Optoelectronics, Huazhong University of Science and Technology, Wuhan 430074, China*
(Dated: January 18, 2022)

**Supplementary Note 1.    MEASUREMENT OF THE QUANTUM FISHER INFORMATION**

### 1.    The QFI and quantum Cramér-Rao bound

In the general quantum parameter estimation experiment, the parameter $\beta$ is usually encoded into a quantum resource state $|\psi(\beta)\rangle$. For a pure state $|\psi(\beta)\rangle$, the quantum Fisher informatio (QFI) is defined as follows

$$\mathcal{F}_\beta = 4\left[\langle\partial_\beta\psi(\beta)|\partial_\beta\psi(\beta)\rangle - |\langle\psi(\beta)|\partial_\beta\psi(\beta)\rangle|^2\right]. \tag{1}$$

The shot-noise limit sensitivity for the parameter estimation by constructing any parameter estimator is bounded by the reciprocal of the square root of the QFI, namely

$$\delta\beta \geq \frac{1}{\sqrt{\mathcal{F}_\beta}}. \tag{2}$$

The is the celebrated quantum Cramér-Rao bound [1, 2].

### 2.    Experimental realization

In the experiment, we utilize the NV center spin in diamond as a two-level quantum sensor to perform a Ramsey interferometry experiment for parameter estimation. The NV center spin is initialized to the spin state $|0\rangle$ and then prepared into the state $|\psi_\theta(0)\rangle = Y_\theta|0\rangle = \cos(\theta/2)|0\rangle - \sin(\theta/2)|-1\rangle$ by an unitary rotation $Y_\theta = \exp(-i\theta\sigma_y/2)$. The free evolution of the system for a time $T$ is governed by the Hamiltonian $H_s = \xi\sigma_z/2$, where $\xi$ represents a magnetic field. This results in the following state that contains the information on the parameter $\beta = \xi T$ as

$$|\psi_\theta(\beta)\rangle = e^{-i\beta H_s}|\psi_\theta(0)\rangle = \begin{bmatrix} \cos(\theta/2)e^{i\beta/2} \\ \sin(\theta/2)e^{-i\beta/2} \end{bmatrix} \tag{3}$$

According to the definition in Eq.(1), the QFI of the state $|\psi_\theta(\beta)\rangle$ with respect to the estimation of the parameter $\beta$ is dependent on the initial resource state $|\psi_\theta(0)\rangle$, namely

$$\mathcal{F}_\beta = \sin^2\theta. \tag{4}$$

In order to measure the QFI of the state $|\psi_\theta(\beta)\rangle$ directly, we first synthesize the microwave driving field using an arbitrary waveform generator

$$f_0(t) = (A\sin\theta)\cos[(\omega_1 - A\cos\theta)t + \beta], \tag{5}$$

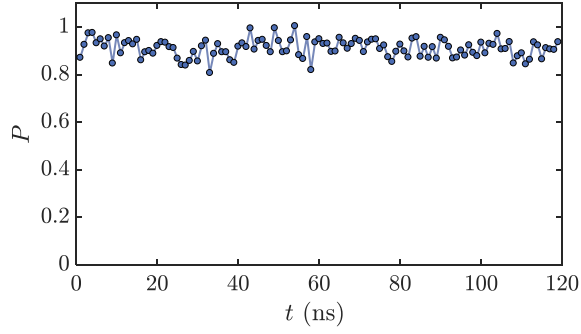

Supplementary Figure 1: Calibration of Hamiltonian engineering. The parameters in the Hamiltonian $\mathcal{H}(\beta)$ (Eq.8) are tuned such that the state $|\psi_\theta(\beta)\rangle$ is approximately its eigenstate. The plot shows the probability of the NV center spin staying in the state $|\psi_\theta(\beta)\rangle$ as a function of the evolution time $t$ when the system's dynamics is governed by the Hamiltonian $\mathcal{H}(\beta)$ (Eq.8). The experimental parameters are $A = (2\pi)15.79$ MHz, $\theta = \pi/2$, $\omega_1 = (2\pi)1440.6$ MHz and $T \simeq 150$ ns.

acting on the NV center spin, which is described by

$$\mathcal{H}(t) = (\omega_1/2)\sigma_z + A\sin\theta\cos\left[(\omega_1 - A\cos\theta)t + \beta\right]\sigma_x. \tag{6}$$

The effective Hamiltonian in a rotating frame with respect to $\mathcal{H}_0 = (1/2)(\omega_1 - A\cos\theta)\sigma_z$ is given by

$$\mathcal{H}(\beta) = e^{i\mathcal{H}_0 t}[\mathcal{H}(t) - \mathcal{H}_0]e^{-iH_0 t} \tag{7}$$

$$= \frac{A}{2}\left(\cos\beta\sin\theta\sigma_x + \sin\beta\sin\theta\sigma_y + \cos\theta\sigma_z\right). \tag{8}$$

The parameters in the above Hamiltonian (Eq.8) are controllable through microwave engineering. In the experiment, we calibrate the above Hamiltonian by verifying that the state $|\psi_\theta(\beta)\rangle$ is approximately its eigenstate, see Supplementary Figure 1. We proceed to implement in our experiment the following time-dependent Hamiltonian by applying the microwave field in the form of Eq.(5) with the designed parametric modulation $\beta \to \beta(t) = \beta + a_\beta \cos(\omega t)$ as

$$\mathcal{H}_{\text{eff}}[\beta(t)] = \mathcal{H}(\beta + a_\beta \cos(\omega t)) \simeq \mathcal{H}(\beta) + a_\beta \cos(\omega t)\partial_\beta \mathcal{H}(\beta). \tag{9}$$

We observe the resonant coherent transition between the eigenstates of the Hamiltonian $\mathcal{H}(\beta)$ (Eq.8) induced by the parameteric modulation, which is shown in Fig.2(b) in the main text. In this case, the parametric modulation frequency $\omega = \omega_0$, where $\omega_0$ is the energy gap between the eigenstates. To measure the state $|\psi_\theta(\beta)\rangle$ population after the parametric modulation for time $\tau$ (the corresponding system's state is denoted as $|\psi_{\theta,\beta}(\tau)\rangle$), we first implement an inverse evolution by $Y_\pi$ and $Y_{\pi-\theta}$ pulses with a free evolution for time $T$ between these two pulses, see Fig.2(a) in the main text. Such an inverse evolution can be described by the following unitary transformation as

$$\hat{U} = Y_{\pi-\theta}\exp\left(-i\xi\beta T\sigma_z\right)Y_\pi \tag{10}$$
$$= Y_\theta^{-1}\exp\left(i\xi\beta T\sigma_z\right) = [\exp\left(-i\xi\beta T\sigma_z\right)Y_\theta]^{-1},$$

which realizes that $\hat{U}|\psi_\theta(\beta)\rangle = |0\rangle$ and $\hat{U}|\psi_\theta^\perp(\beta)\rangle = |-1\rangle$. The subsequent spin-dependent fluorescence measurement $\hat{P}_0 = |0\rangle\langle 0|$ is thus equivalent to the projective measurement $\hat{P} = \hat{U}^\dagger|0\rangle\langle 0|\hat{U} = |\psi_\theta(\beta)\rangle\langle\psi_\theta(\beta)|$ on the state $|\psi_{\theta,\beta}(\tau)\rangle$. Therefore, we are able to monitor the state $|\psi_\theta(\beta)\rangle$ population dynamics under resonant parametric modulation which can be described by $P_\beta(t) = [1 + \cos(\nu_\theta t)]/2$, where [3, 4]

$$\nu_\theta = (1/2)a_\beta\omega_0\sqrt{\mathcal{F}_\beta[\psi_\theta(\beta)]}. \tag{11}$$

Therefore, we are able to determine the QFI of the state $|\psi(\beta)\rangle$ as follows

$$\mathcal{F}_\beta[\psi_\theta(\beta)] = 4\left(\frac{\nu_\theta}{a_\beta\omega_0}\right)^2. \tag{12}$$

# Supplementary Note 2. VERIFICATION OF QUANTUM CRAMÉR-RAO BOUND

## 1. Parameter estimation via Ramsey interferometry experiment

In our experiment, we perform quantum parameter estimation based on Ramsey interferometry, as shown in Fig.1(a) in the main text. The two-level quantum system is realized by the spin sublevels in the ground state manifold of the NV center, i.e. $|m_s = 0\rangle$ and $|m_s = -1\rangle$. The system is coherently manipulated by microwave field pulses which are described by the following Hamiltonian

$$H_{pulse} = (\omega_1/2)\sigma_z + A\sin[(\omega_1 + \delta)t]\sigma_x, \tag{13}$$

for $t \in [t_0, t_0 + \tau_{pulse})$, where $\omega$ denotes the energy gap between the states $|0\rangle$ and $|-1\rangle$ when applying microwave field, and $\tau_{pulse}$ represents the time duration for microwave pulse. We remark that the energy splitting of the spin sublevels may slightly change due to microwave driving. In the interaction picture with respect to $H_0 = (\omega_1 + \delta)/2\sigma_z$, we get the following effective Hamiltonians during the microwave pulses ($H_I^{(1)}$) and the free evolution ($H_I^{(0)}$) respectively

$$H_I^{(1)} = -\frac{\delta}{2}\sigma_z + \frac{A}{2}\sigma_y \tag{14}$$

$$H_I^{(0)} = -\frac{\xi}{2}\sigma_z \tag{15}$$

with $\xi = \delta + (\omega_1 - \omega_0)$ where $\omega_0$ denotes the energy gap between the states $|0\rangle$ and $|-1\rangle$ during the free evolution.

In the experiment, the NV center spin is initialized to $|0\rangle$ and then is prepared into the $\theta$-dependent resource state $|\psi_\theta(0)\rangle$ by an unitary rotation $Y_\theta = \exp(-i\theta\sigma_y/2)$ which is realized by applying a microwave pulse with a Rabi frequency $\Omega$ for a time duration $\tau_{pulse_1} = \theta/\Omega$. Here, we remark that $\delta \ll \Omega$, thus the error in the rotation is negligible. The free evolution process (from $\tau_{pulse_1}$ to $\tau_{pulse_1} + T$) leads to a dynamical phase accumulation given by the parameter $\beta = \xi T$, and the system evolves to the following final state as

$$\begin{aligned}|\psi(\theta,\beta)\rangle &= \exp(i\xi T\sigma_z/2)Y_\theta|0\rangle \\ &= \cos(\theta/2)e^{-i\beta/2}|0\rangle - \sin(\theta/2)e^{i\beta/2}|-1\rangle.\end{aligned} \tag{16}$$

The second unitary rotation $Y_\alpha = \exp(i\alpha\sigma_y/2)$ to implement the projective measurement $\hat{P}_\alpha$, is realized by a microwave pulse for a time duration $\tau_{pulse_2} = \alpha/\Omega$, see Supplementary Figure 2, implements the projective measurement $\hat{P}_\alpha = |\phi_\alpha\rangle\langle\phi_\alpha|$ where $|\phi_\alpha\rangle = \cos(\alpha/2)|0\rangle + \sin(\alpha/2)|-1\rangle$. In the experiment, we choose the free evolution time $T$ such that the working point is close to $\beta \simeq \pi/2 + k\pi$ where the measurement signal exhibits the maximum slope.

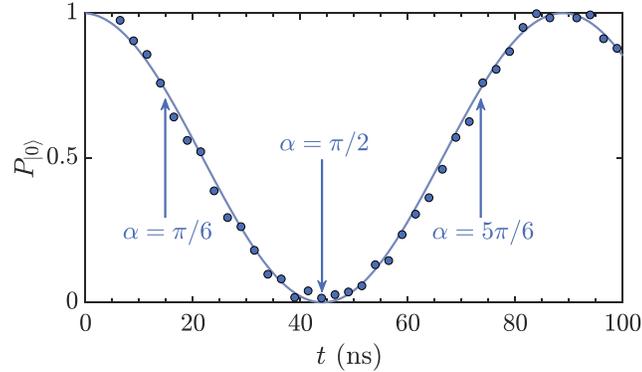

Supplementary Figure 2: Microwave pulse for the implementation of projective measurement. The projective measurement $\hat{P}_\alpha = |\phi_\alpha\rangle\langle\phi_\alpha|$, where $|\phi_\alpha\rangle = \cos(\alpha/2)|0\rangle + \sin(\alpha/2)|-1\rangle$, can be realized by an unitary rotation $Y_\alpha = \exp(-i\alpha\sigma_y/2)$ before the spin-dependent fluorescence measurement. The red dots show the population of the state $|0\rangle$ as a function of the microwave pulse duration, which allows to determine Rabi period $T_{Rabi}$. The rotation $Y_\alpha$ can be realized by setting the microwave pulse duration time as $\tau_\alpha = \alpha T_\alpha/\pi$. As an example, we mark three microwave pulse duration times $\tau_\alpha$ for $\alpha = \pi/6, \pi/2, 5\pi/6$ in the figure.

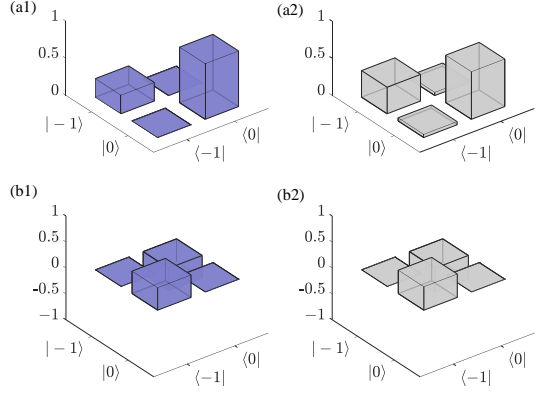

Supplementary Figure 3: State preparation fidelity. We plot the density matrix elements of the prepared state $\rho$ based on the results from a set of different projective measurements $\hat{P}_\alpha$, which are compared with the ones of the ideal state $|\psi(\theta,\beta)\rangle\langle\psi(\theta,\beta)|$. The four panels show the real (a1, a2) and imaginary (b1, b2) parts of the density matrices, wherein the blue bars show the ideal state and the solid gray bars show the density matrix of the estimated prepared state. The fidelity is estimated to be $F = 96.2\%$, which is defined as $F = \langle\psi(\theta,\beta)|\rho(\beta;\theta)|\psi(\theta,\beta)\rangle$. The parameters are $\theta = \pi/3$ and $\beta = \pi/2$.

In additional, we reconstruct the prepared state according to the measurement results $p(\beta;\theta,\alpha) = \text{Tr}\{\rho(\beta;\theta)\hat{P}_\alpha\}$ from a set of 11 different projective measurements $\hat{P}_\alpha$. By performing the following minimization procedure as

$$\min_{\{r,\theta_e,\phi_e\}}\left\{\sum_\alpha \left[\text{Tr}\{\rho^R(r,\theta_e,\phi_e)\hat{P}_\alpha\} - p(\beta;\theta,\alpha)\right]^2\right\}, \quad (17)$$

wherein $\rho^R(r,\theta_e,\phi_e) = 1/2[\mathbb{1} + r(\sin\theta_e\cos\phi_e\sigma_x + \sin\theta_e\sin\phi_e\sigma_x + \cos\theta_e\sigma_z)]$, we can get the most likely density matrix $\rho$ for the prepared state. As an example, our estimation suggests a state preparation fidelity of $F = 96.2\%$ fidelity in Supplementary Figure 3 with $F = \langle\psi(\theta,\beta)|\rho(\beta;\theta)|\psi(\theta,\beta)\rangle$.

## 2. Quantum parameter estimation protocol

The sensitivity of quantum parameter estimation is dependent on the measurement protocol. In the experiment, we perform projective measurement on the NV center spin that is described by the operator $\hat{P}_\alpha = |\phi_\alpha\rangle\langle\phi_\alpha|$ with the basis state $|\phi_\alpha\rangle = \cos(\alpha/2)|0\rangle + \sin(\alpha/2)|-1\rangle$. We count the number of photons in the first 300 ns of the laser pulse as the signal photons. Due to the limit of collection efficiency, the signal photons are accumulated over a number of sweeps of an experimental measurement sequence, which constitutes one experiment run of measurement. We denote the averaged photon number obtained from the bare spin state $m_s = 0$ and $m_s = -1$ as $n_0$ and $n_1$ respectively. We introduce a variable $s = 1/0$ to represent the spin state $m_s = 0/m_s = -1$. For the NV center spin system, the signal photons are spin-dependent, namely $(n_0 - n_1)/n_0 \simeq 30\%$, see Supplementary Figure 4(a). For a quantum state $\rho$ with the state $|0\rangle$ population $p = \langle 0|\rho|0\rangle$, the number of photons $n_j$ collected in the $j$-th experiment run fluctuates and follows the distribution $n_j \sim p\mathcal{N}(n_0,\sigma_0^2) + (1-p)\mathcal{N}(n_1,\sigma_1^2)$, see an example shown in Supplementary Figure 4(b). According to the properties of the normal distribution[11], the random variable $p_j = (n_j - n_1)/(n_0 - n_1)$ follows the probability distribution $Q(p)$

$$p_j \sim Q(p_j) = p\mathcal{N}(1,\tilde{\sigma}_0^2) + (1-p)\mathcal{N}(0,\tilde{\sigma}_1^2), = \mathcal{N}(p, p^2\tilde{\sigma}_0^2 + (1-p)^2\tilde{\sigma}_1^2),$$

where $\Delta n = n_0 - n_1$ and $\tilde{\sigma}_m = \sigma_m/\Delta n$, $m = 0,1$. $Q(p_j)$ is shown in Supplementary Figure 5 and is divided into a series of intervals by the integers $k = \lfloor p_j \rfloor$.

Based on the distribution $Q(p_j)$, we proceed to assign a measurement value $s_j = k + 1$ or $k$ according to the probabilities $p_j^{(k)} = p_j - k$ and $1 - p_j^{(k)}$ in the $k^{\text{th}}$ interval. This allows us to construct a quantity as $S = (1/N)\sum_{j=1}^N s_j$,

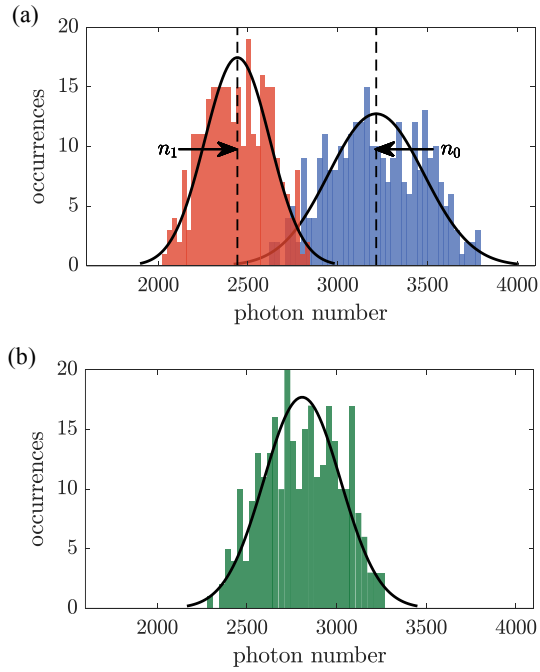

Supplementary Figure 4: (a) shows the histogram of the number of photons collected from the spin state $|-1\rangle$ (red) and $|0\rangle$ (blue) with the averaged number of photons $n_1$ and $n_0$ respectively. (b) shows the histogram of the number of photons collected while the NV center spin is in the superposition state $|+\rangle = (1/\sqrt{2})(|0\rangle + |-1\rangle)$.

the expectation value of which is

$$\langle S \rangle = \frac{1}{N} \langle \sum_{j=1}^{N} s_j \rangle \tag{18}$$

$$= \frac{1}{N} \sum_{j=1}^{N} \left\{ \sum_k \int_k^{k+1} dp_j \left[ (k+1)(p_j - k)Q(p_j) + k(1 - p_j + k)Q(p_j) \right] \right\} \tag{19}$$

$$= \frac{1}{N} \sum_{j=1}^{N} \left\{ \sum_k \int_k^{k+1} dp_j p_j Q(p_j) \right\} = p. \tag{20}$$

The variance of the quantity $S$ is given by

$$(\Delta s)^2 = \langle S^2 \rangle - \langle S \rangle^2 \tag{21}$$

$$= \langle [\frac{1}{N} \sum_j^N (s_j - p)]^2 \rangle \tag{22}$$

$$= \frac{1}{N^2} \left\langle \sum_{j=1}^N s_j^2 + \sum_{j \neq k}^N s_j s_k - 2Np \sum_{j=1}^N s_k + N^2 p^2 \right\rangle \tag{23}$$

$$= \frac{1}{N} \left( \langle s_j^2 \rangle - p^2 \right) \tag{24}$$

$$\tag{25}$$

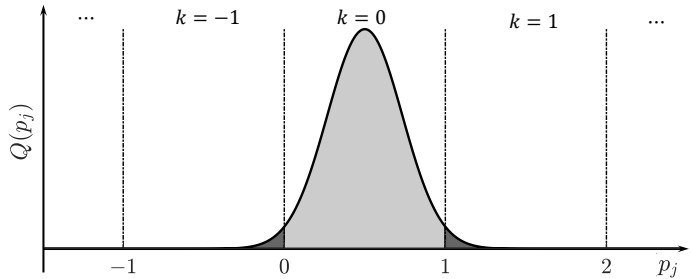

Supplementary Figure 5: The probability distribution $Q(p_j)$ is divided into a series of interval by $k = \lfloor p_j \rfloor$.

The first term can be calculated as

$$\langle s_j^2 \rangle = \sum_k \int_k^{k+1} dp_j \left[ (k+1)^2 (p_j - k) Q(p) + k^2 (1 - p_j + k) Q(p) \right] \tag{26}$$

$$= \sum_k \int_k^{k+1} dp_j Q(p) [(2k+1) p_j - k(k+1)] \tag{27}$$

$$= \sum_k \int_k^{k+1} dp_j \left[ p_j Q(p) \right] + \sum_k \int_k^{k+1} dp_j \left[ k(2p_j - k - 1) Q(p) \right] \tag{28}$$

$$= p + \sum_k F_k. \tag{29}$$

We define $p_j = \lfloor p_j \rfloor + \delta_j = k + \delta_j$ with $\delta_j \in [0, 1)$, and $F_k$ can be write as

$$F_k = \int_k^{k+1} dp_j k(k - 1 + 2\delta_j) Q(p). \tag{30}$$

It can be seen that for all $k \in Z$, $F_k \geq 0$, and if and only if $k = 0$, $F_k = 0$. Therefore, Eq.(29) satisfies

$$\langle s_j^2 \rangle \geq p, \tag{31}$$

and the variance of the quantity $S$ is bounded by the shot noise

$$(\Delta s)^2 \geq \frac{1}{N} (1 - p) p. \tag{32}$$

If the distribution $Q(p)$ is strictly localized in the zeroth ($k = 0$) interval, i.e the black areas in Supplementary Figure 5 are negligible, all the components $F_k \simeq 0$. Therefore, the variance of the observable $S$ achieves the shot noise

$$(\Delta s)^2 = \frac{1}{N} (1 - p) p. \tag{33}$$

In our experiment, the measurements are performed at the working points $\beta = \pi/2$, which makes $p \simeq 1/2$ and

$$Q(p_j) \simeq \mathcal{N}(1/2, \sigma^2), \tag{34}$$

with $\sigma = (1/2) \sqrt{\tilde{\sigma}_0^2 + \tilde{\sigma}_1^2}$. The distribution of $p_j$ obtained in our experiment satisfies $\sqrt{\tilde{\sigma}_0^2 + \tilde{\sigma}_1^2} \simeq 1/2$, which guarantees a more than 95% confidence interval of $k = 0$ (see Supplementary Figure 5).

Furthermore, we note that

$$p = \frac{1}{2} \left( 1 + \cos \theta \cos \alpha - \sin \theta \sin \alpha \cos \beta \right), \tag{35}$$

thus we can construct the following estimator for the parameter $\beta$ as

$$\check{\beta} = \arccos \left( \frac{1 + \cos \theta \cos \alpha - 2S}{\sin \theta \sin \alpha} \right) \tag{36}$$

$$= \arccos \left[ \frac{1}{\sin \theta \sin \alpha} \left( \cos \theta \cos \alpha - \frac{N_0 - N_1}{N} \right) \right],$$

where $N_0$ and $N_1$ represents the number of $s_j = 0$ and 1 respectively. With $\alpha = \pi/2$, the estimator becomes

$$\check{\beta} = \arccos\left[\frac{1}{\sin\theta}\left(\frac{N_1 - N_0}{N}\right)\right]. \tag{37}$$

The precision can be written as

$$\delta\beta = \frac{\Delta s}{\left|\partial_{\langle\hat{S}\rangle}\check{\beta}\right|} = \frac{2\Delta s}{|\sin\theta\sin\alpha\sin\beta|} \tag{38}$$

which gives the optimal sensitivity with $\alpha = \pi/2$ satisfying the quantum Cramér-Rao bound.

### 3. Optimal measurement to achieve quantum Cramér-Rao bound

In our experiment, we perform projective measurement on the state $|\psi_\theta(\beta)\rangle = \cos(\theta/2)e^{-i\beta/2}|0\rangle - \sin(\theta/2)e^{i\beta/2}|-1\rangle$ to estimate the value of the parameter $\beta$. We compare the measurement sensitivity achieved by different projective measurements, which are described by $\hat{P}_\alpha = |\phi_\alpha\rangle\langle\phi_\alpha|$ with $|\phi_\alpha\rangle = \cos(\alpha/2)|0\rangle + \sin(\alpha/2)|-1\rangle$. The measurement signal obtained from different projective measurements are shown in Supplementary Figure 6. Following the protocol as presented in the above section, we analyze the variance of parameter estimation and thereby obtain the measurement sensitivity from the projective measurement $\hat{P}_\alpha$. We find that the optimal sensitivity is obtained by the projective measurement $\hat{P}_{\pi/2} = |+\rangle\langle+|$ with $|+\rangle = \frac{1}{\sqrt{2}}(|0\rangle + |-1\rangle)$, where

$$\langle\hat{P}_{\pi/2}\rangle = \frac{1}{2}(1 - \cos\beta\sin\theta) \tag{39}$$
$$\Delta\hat{P}_{\pi/2} = \sqrt{\langle\hat{P}_{\pi/2}^2\rangle - \langle\hat{P}_{\pi/2}\rangle^2} = \frac{1}{2}\sqrt{1 - (\cos\beta\sin\theta)^2},$$

which gives the optimal measurement sensitivity as follows

$$\delta\beta = \frac{\sqrt{1 - (\cos\beta\sin\theta)^2}}{|\sin\beta\sin\theta|}. \tag{40}$$

We set the free evolution time such that the parameter $\beta = \Delta T$ is close to the working point with the maximum slope of the measurement signal, namely $\beta \simeq (k \pm 1/2)\pi$. In this case, the optimal measurement sensitivity (Eq.40) can be written as

$$\delta\beta|_{\beta=\frac{\pi}{2}} = \sin^{-1}\theta, \tag{41}$$

which equals to $1/\sqrt{\mathcal{F}_\beta}$. We note that the QFI is $\mathcal{F}_\beta = \sin^2\theta$. Therefore, by the projective measurement $\hat{P}_{\pi/2} = |+\rangle\langle+|$ with $|+\rangle = \frac{1}{\sqrt{2}}(|0\rangle + |-1\rangle)$ we achieve the sensitivity limit and verify its connection with quantum Cramér-Rao bound, see Fig.3(d) in the main text.

### Supplementary Note 3. THE QFI AND QUANTUM ENTANGLEMENT

#### 1. The interacting Hamiltonian of a correlated two-qubit system

In this section, we investigate in detail the feasibility of the parametric modulation scheme to measure the QFI in a correlated two-qubit system which has been utilized to experimentally extract the complete quantum geometric tensor [4]. The system is formed by the NV center spin in diamond and a nearby $^{13}$C nuclear spin, which is described by the following Hamiltonian as

$$\mathcal{H} = D_{gs}S_z^2 + \gamma_e B_z S_z + \gamma_C B_z I_z + A_\parallel S_z \otimes I_z + A_\perp S_z \otimes I_x, \tag{42}$$

where $S$ and $I$ are the spin operators of NV center spin and $^{13}$C nuclear spin. The zero-field splitting is $D_{gs} = (2\pi)2.87$GHz, $\gamma_e = (2\pi)2.8$ MHz/G and $\gamma_C = (2\pi)1.07$ kHz/G are the electronic spin and nuclear spin gyromagnetic

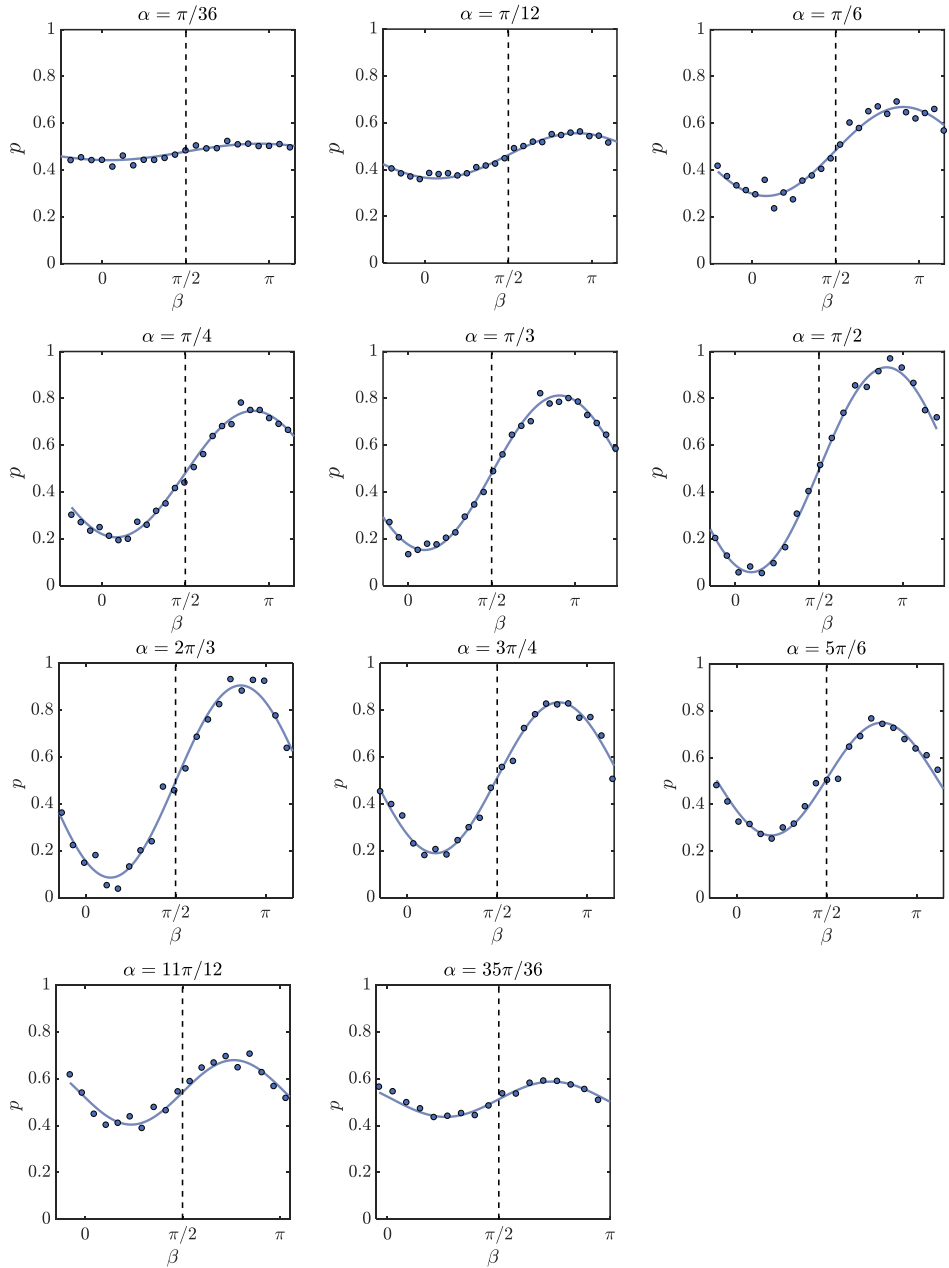

Supplementary Figure 6: Parameter estimation with different projective measurements. The measurement signal $p = \langle \hat{P}_\alpha \rangle$ from the projective measurement $\hat{P}_\alpha = |\phi_\alpha\rangle\langle\phi_\alpha|$ with $|\phi_\alpha\rangle = \cos(\alpha/2)|0\rangle + \sin(\alpha/2)|-1\rangle$ is shown as a function of the phase parameter $\beta$. The projective measurement $\hat{P}_{\pi/2}$ leads to the maximum signal contrast, see panel (f), which enables us to achieve the optimum measurement sensitivity for the estimation of the parameter $\beta$. The parameter is $\theta = \pi/2$.

ratio, respectively. The experimentally determined hyperfine coupling parameters are $A_\perp = (2\pi)2.79$ MHz, $A_\parallel = (2\pi)11.832$ MHz [4], and the external magnetic field is $B_z = 504$ G.

In our scheme, the first qubit is formed by the spin sublevels $m_s = -1$ and $m_s = 0$ of the NV center electronic spin as $|0\rangle \equiv |m_s = -1\rangle$ and $|1\rangle \equiv |m_s = 0\rangle$. The second qubit is encoded on $^{13}$C nuclear spin as $|0\rangle \equiv |+1/2\rangle$ and

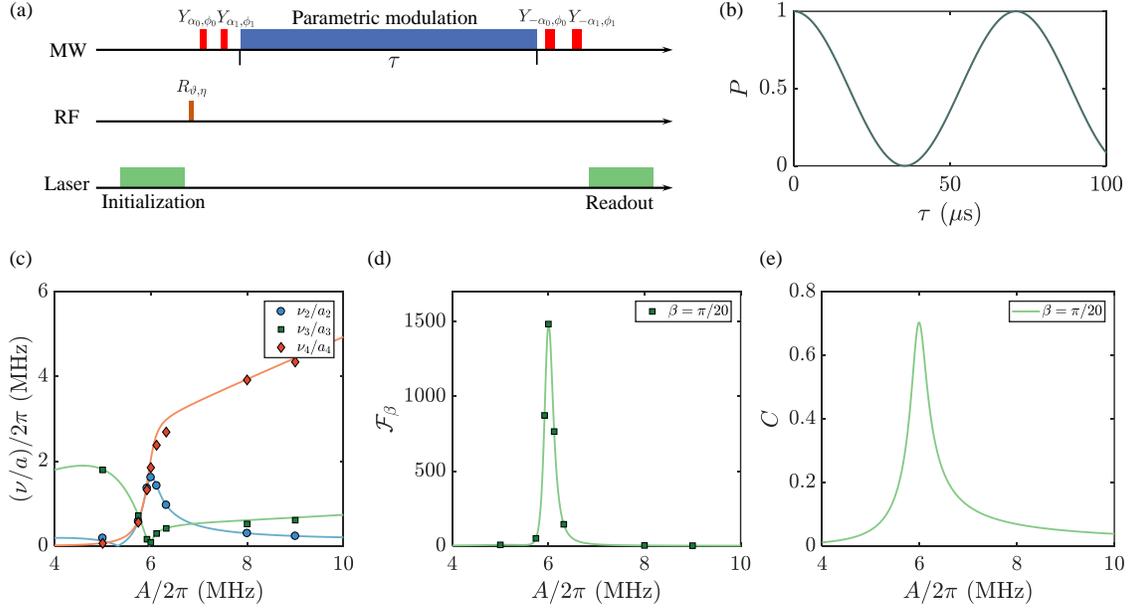

Supplementary Figure 7: **The QFI and quantum entanglement in a correlated two-qubit system**. (a) The pulse sequence for the measurement of the QFI. The state preparation is achieved by applying three pulses $R_{\vartheta,\eta}$, $Y_{\alpha_0,\phi_0}$ and $Y_{\alpha_1,\phi_1}$ successively after the optical initialization by a green laser (532 nm) pulse. A subsequent parametric modulation microwave field is applied to induce the transition between the state $|\Psi_1\rangle$ and the other eigenstates. The effective Rabi frequency can be extracted from the population dynamics of the NV center spin state $|m_s = 0\rangle$ after applying two inverse pulses $Y_{-\alpha_0,\phi_0}$ and $Y_{-\alpha_1,\phi_1}$. (b) The coherent oscillation of the population of the NV center spin state $|m_s = 0\rangle$ as a function of time, which indicates the transition between $|\Psi_1\rangle$ and $|\Psi_3\rangle$ with $A = (2\pi)3$ MHz. (c) The three Rabi frequencies of the transition induced by parametric modulation as a function of the parameter $A$. (d) The QFI $\mathcal{F}_\beta$ of the ground state $|\Psi_1\rangle$ as a function of the parameter $A$. The data points in (c-d) are obtained from the simulation of the experiments which agree well with the exact theoretical values (solid lines). (e) The concurrence $C$ of the state $|\Psi_1\rangle$ as a function of the parameter $A$. The other parameters we use in (c-e) are $A_\perp = (2\pi)2.79$ MHz, $A_\parallel = (2\pi)11.832$ MHz, $B_z = 504$ G.

$|1\rangle \equiv |-1/2\rangle$. The Hamiltonian describing the correlated two-qubit system can be rewritten as

$$\mathcal{H} = \frac{\omega_1}{2}\sigma_z + \left(\frac{\gamma_C B_z}{2} - \frac{A_\parallel}{4}\right)\tau_z - \frac{A_\perp}{4}\tau_x$$
$$- \frac{A_\parallel}{4}\sigma_z \otimes \tau_z - \frac{A_\perp}{4}\sigma_z \otimes \tau_x, \tag{43}$$

wherein $\sigma$ and $\tau$ are the Pauli matrices for two qubits and $\omega_1 = D_{gs} - \gamma_e B_z$ denotes the energy gap between the NV center spin sublevels $m_s = 0, -1$. A microwave driving field is applied on the system

$$\mathcal{H}_{mw} = A \sin\beta_t \cos\left[\omega_1 t + 2A \int_0^t \cos\beta_\tau d\tau + \varphi\right] \sigma_x. \tag{44}$$

The effective Hamiltonian in a rotating frame is given by

$$\mathcal{H}(\beta) = e^{i\mathcal{K}(t)}\mathcal{H}(t)e^{-i\mathcal{K}(t)} + i\left(\frac{\partial e^{i\mathcal{K}(t)}}{\partial t}\right)e^{-i\mathcal{K}(t)} \tag{45}$$
$$= \frac{A}{2}[\cos\beta\sigma_z + \sin\beta(\cos\varphi\sigma_x + \sin\varphi\sigma_y)] - \frac{A_\parallel}{4}\sigma_z\tau_z$$
$$- \frac{A_\perp}{4}\sigma_z\tau_x + \left(\frac{\gamma_C B_z}{2} - \frac{A_\parallel}{4}\right)\tau_z - \frac{A_\perp}{4}\tau_x,$$

with the operator $\mathcal{K}(t)$ defined as

$$\mathcal{K}(t) = \left(\frac{\omega_1}{2}t + A\int_0^t \cos\beta_\tau d\tau\right)\sigma_z. \tag{46}$$

We denote the four eigenstates of the Hamiltonian as $|\Psi_1\rangle, |\Psi_2\rangle, |\Psi_3\rangle, |\Psi_4\rangle$ with their associated eigenvalues $\epsilon_1 < \epsilon_2 < \epsilon_3 < \epsilon_4$.

## 2. Simulation of the QFI measurement and quantum entanglement

The experiment scheme that we simulate is shown in Supplementary Figure 7(a). The system can be initialized into the state

$$\begin{aligned}|\Psi_1\rangle &= \sin\vartheta e^{i\eta}\left(\sin\alpha_0 e^{i\phi_0}|0\rangle + \cos\alpha_0|1\rangle\right) \otimes |0\rangle \\ &+ \cos\vartheta \left(\sin\alpha_1 e^{i\phi_1}|0\rangle + \cos\alpha_1|1\rangle\right) \otimes |1\rangle,\end{aligned} \qquad (47)$$

by applying a radio frequency (RF) pulse $R_{\vartheta,\eta} = \exp[i\vartheta(\sin\eta\tau_x + \cos\eta\tau_y)]$ conditioning on the NV center spin state $|1\rangle = |m_s = 0\rangle$, and two selective microwave pulses $Y_{\alpha_0,\phi_0} = \exp[i\alpha_0(\sin\phi_0 X_0 + \cos\phi_0 Y_0)]$ and $Y_{\alpha_1,\phi_1} = \exp[i\alpha_1(\sin\phi_1 X_1 + \cos\phi_1 Y_1)]$ with $X_j = \sigma_x \otimes |j\rangle\langle j|$ and $Y_j = \sigma_y \otimes |j\rangle\langle j|$ ($j = 0, 1$). We remark that although we prepare the parameter-dependent state via coherent pulse control, it is also feasible to prepare the system into the eigenstate via adiabatic evolution.

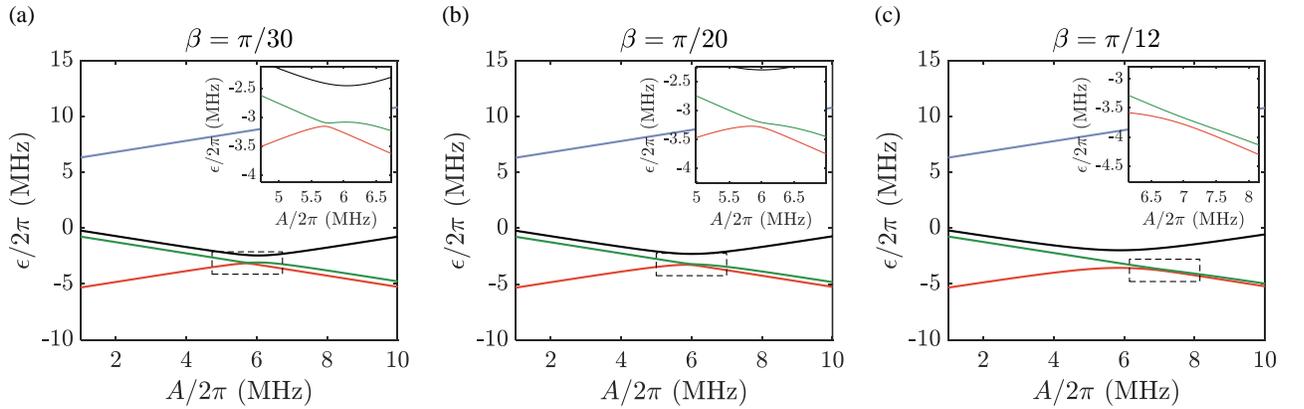

Supplementary Figure 8: The eigenvalues $\epsilon$ of the Hamiltonian $\mathcal{H}(\beta)$ with $\beta = \pi/30, \pi/20, \pi/12$. The other parameters are the same as Fig.4 in the main text.

The QFI associated with the ground state $|\Psi_1\rangle$ can be extracted by applying the parametric modulation $\beta_t = \beta + a_k\cos(\omega_k t)$ for $k \in \{2, 3, 4\}$. We simulate the parametric modulation induced Rabi oscillation between the ground state $|\Psi_1\rangle$ and the other three eigenstates, see e.g. Supplementary Figure 7(b), from which one can determine the QFI as follows

$$\mathcal{F}_\beta = 4\sum_{k=2}^{4}\left(\frac{\nu_k}{a_k\omega_k}\right)^2, \qquad (48)$$

where $\nu_k$ is the corresponding Rabi frequency under the resonant condition $\omega_k \simeq \epsilon_k - \epsilon_1$, as shown in Supplementary Figure 7(c-d). Furthermore, we calculate the concurrence [5] of the ground state $|\Psi_1\rangle$, see Supplementary Figure 7(d), which shows that the system exhibits a high level of entanglement.

In the main text (Fig. 4) and Supplementary Figure 7, we show that the QFI of the ground state reaches a very large value. We plot the energy spectrum of the system, see Supplementary Figure 8. It can be seen that the peak value of the QFI appears when the energy gap is small, i.e. the anti-crossing point. In this case, the ground state $|\Psi_1\rangle_\beta$ becomes very sensitive to the parameter $\beta$. This can be qualitatively understood in the following way. According to the perturbation theory, the ground state $|\Psi_1\rangle_{\beta+\delta\beta}$ can be expressed as

$$|\Psi_1\rangle_{\beta+\delta\beta} = |\Psi_1\rangle_\beta + \delta\beta\sum_{k=2}^{4}\frac{\langle\Psi_i|\partial_\beta\mathcal{H}(\beta)|\Psi_1\rangle}{E_1 - E_k}|\Psi_k\rangle. \qquad (49)$$

Therefore, the small energy gap would result in a significant state derivation $|\partial_\beta \Psi_1\rangle$, which indicates a very large value of QFI (see Eq.(4) in the main text). In addition, we also note that the level anticrossing is usually an evidence for ground state entanglement in interacting systems [6, 7]. This explains the prominent entanglement which is shown in Fig.4 (b) in the main text. As the level anticrossing may exist in a variety of interacting many-body systems, the connection between the QFI and entanglement revealed by the present example is expected to be a general many-body phenomenon.

### 3. Extension to many-body quantum systems

We note that many-body quantum systems represent valuable resources for quantum metrology. To exploit the present scheme for the direct measurement of the QFI in many-body quantum systems, it requires that the parameter-dependent resource state shall be the ground state of a certain Hamiltonian (namely parent Hamiltonian). Although it is not straightforward to present the parent Hamiltonian for any many-body resource state, the formalism of matrix product state (more general projected entangled pair state and tensor network state [8]) provide a systematic way to find the parent Hamiltonian. It has been proven that matrix product states can efficiently describe many-body quantum states, and a parent Hamiltonian can be constructed for any matrix product state [9]. The other important ingredient for the measurement of the QFI relies on the system's dynamical response under parametric modulation. In the scenario of many-body systems, it would be more efficient to measure the total excitation rate following the Fermi-Golden-rule approach in Ref. [3]. We remark that the present technique will be of particular interest for quantum metrology of estimating a general parameter of a Hamiltonian [10] which would be important to explore the metrological potential of many-body quantum systems.



\end{document}